\documentclass[12pt]{article}
\newcommand{\be}{\begin{equation}}
\newcommand{\ee}{\end{equation}}
\newcommand{\bea}{\begin{eqnarray}}
\newcommand{\eea}{\end{eqnarray}}
\newcommand{\nn}{\nonumber}
\newcommand{\ql}{\mathcal{L}}
\newcommand{\hf}{{1\over 2}}
\newcommand{\of}{{1\over 4}}
\newcommand{\p}{\partial}
\newcommand{\ba}{\bar{A}}
\newcommand{\ua}{\hat{A}}
\newcommand{\bF}{\bar{F}}
\newcommand{\uF}{\hat{F}}
\newcommand{\bp}{\bar{p}}
\newcommand{\up}{\hat{p}}
\newcommand{\bH}{\mathcal{H}}

\begin{document}
\begin{titlepage}

\title{Hamiltonian analysis of the 
proposal by Chen et. al., PRL 100 (2008) 232002}

\author{M.N. Stoilov\\
{\small\it Bulgarian Academy of Sciences,}\\
{\small\it Institute of Nuclear Research and Nuclear Energy,}\\
{\small\it Blvd. Tzarigradsko Chausse\'e 72, Sofia 1784, Bulgaria}\\
{\small e-mail: mstoilov@inrne.bas.bg}}

\maketitle

\begin{abstract}
It is proposed by Chen et. al.  to represent the gauge fields in theories with local symmetries as a sum of "physical" and "pure gauge"  fields which to be treated separately. Here we show that after quantization this representation  leads  to a model which is non equivalent to the initial one.
\end{abstract}


key words: Electrodynamics, gauge invariance, constraint systems

\end{titlepage}
\section{Introduction}
In a series of  recent papers \cite{c1, c2}  a decomposition of the  gauge field in models with local symmetry into a physical field  and a pure gauge one is proposed.
The same scheme is applied later to the gravitational field as well \cite{c3}.
The proposed procedure has significant consequences, e.g. in Quantum Chromodynamics, but also rise a lot of questions \cite{t1, j1, j2}.
Here we want to consider the problem from the viewpoint of Hamiltonian system with constraints  and to apply to it the Dirac procedure for quantization \cite{ht}. 
It is enough for our purposes to investigate the simplest case of free  Electromagnetic ($U(1)$ gauge) field, because the specificity of the treatment is same both for Abelian and non Abelian gauge fields.
The interaction with matter also does not change the considerations.

\section{Decomposition of the $U(1)$ gauge field}
The basic idea in Refs.\cite{c1, c2, c3} is to represent  the gauge field $A$ as a sum of two fields --- the "physical" field $\ua$ and the "pure gauge" field $\ba$
\be 
A=\ua+ \ba.\label{decom}
\ee
The conditions, which are proposed to distinguish the nature of these fields in the case of 
$U(1)$ local symmetry are
\bea 
\p_i \ua_i&=&0, \label{zx1}\\ 
\bF_{\mu\nu}\equiv\p_\mu \ba_\nu-\p_\nu \ba_\mu &=&0.\label{zx2}
\eea

Eq.(\ref{decom}) effectively doubles the gauge potential, while eqs.(\ref{zx1},\ref{zx2}) are used to decrease back the number of the degrees of freedom.
In this sense, among eqs.(\ref{decom}--\ref{zx2}) the really important one is eq.(\ref{decom}).
This is the reason to start our analysis considering the consequences of eq.(\ref{decom}) itself, i.e., for a moment we treat $\ua$ and $\ba$ as  independent fields.

The pure Electromagnetic Lagrangean is
\be
L=- \of F_{\mu\nu}F_{\mu\nu}, \label{la1}
 \ee
where $F_{\mu\nu}=\p_\mu A_\nu-\p_\nu A_\mu$.
Substituting eq.(\ref{decom}) into eq.(\ref{la1}) we obtain:
\be 
\ql =- \of \uF_{\mu\nu}\uF^{\mu\nu} - \of \bF_{\mu\nu}\bF^{\mu\nu} 
- \hf \uF_{\mu\nu}\bF_{\mu\nu} \label{la2}
\ee
($\uF_{\mu\nu}=\p_\mu \ua_\nu-\p_\nu \ua_\mu $).
The Lagrangean(\ref{la2}) possesses an enlarged gauge symmetry.
It is  not only $U(1)$ invariant but is also invariant under the so called Stuckelberg gauge symmetry:
\bea
\delta \ua&=& C \nn \\
\delta \ba&=& -C. \label{sg}
\eea
Here $C$ is an arbitrary real $4-$vector field.
The existence of the gauge transformation (\ref{sg}) allows us to prove that
 the models with Lagrangean  (\ref{la2}) and  (\ref{la1}) are equivalent.

Proof:

Perform the following change of variables 
\bea 
A&=& \ua + \ba\\
B&=& \hf(\ua -\ba)
\eea
The Jacobean of this transformation is $1$ and the Lagrangean (\ref{la2}) does not depend on $B$.
Therefore, the continual integration over $B$ field is trivial and the transition amplitude for the model with Lagrangean (\ref{la2}) coincides with the transition amplitude for the pure quantum Electromagnetic field.

A more canonical approach to the proof of the equivalence involves the use of the  gauge fixing 
\be B_\mu=0. \label{sgf}\ee
The Faddeev--Popov determinant which corresponds to eqs.(\ref{sg},\ref{sgf})  is $1$.
Therefore, there are no ghosts, the gauge conditions (\ref{sgf}) can be used to trivially integrate over the $B$ field and thus the equivalence is proved.

The above considerations demonstrate that the Lagrangean (\ref{la2}) is just an uneconomic  way to describe a well known model.
Its potential advantage compare to the Lagrangean (\ref{la1}) is if, eventually, we find a gauge in which the fields $\ua$ and $\ba$ are the physical and pure gauge part of the Electromagnetic field. 
In other words, the question is whether eqs.(\ref{zx1},\ref{zx2}) can be used as gauge conditions for the Stuckelberg gauge transformation (\ref{sg}).

\section{Hamiltonian analysis}
 Let $\up$ and $\bp$ are  the  momenta conjugate to the fields $\ua$ and $\ba$.
 From the Lagrangean (\ref{la2}) we obtain the following primary constraints:
 \bea 
 \up_0 &=& 0 \label{pci}\\
 \bp_0 &=& 0 \label{pcii}\\
\up_i-\bp_i &=& 0 \label{pcf}
\eea
 and the following secondary constraints:
\bea 
\p_i \up_i &=& 0 \label{sci}\\
\p_i \bp_i  &=& 0. \label{scf}
\eea
Constraints (\ref{pci},\ref{pcii}) indicates that the corresponding dynamically conjugated variables $\ua_0$ and $\ba_0$ are Lagrange multipliers.
It is also clear  that not all constraints (\ref{pcf}--\ref{scf})  are independent.
So, we have to choose a subset of them and 
we use the freedom in this choice  to achieve 
our primary goal --- to make $\ua$  the physical part of the Electromagnetic field $A$, i.e., to make this field $U(1)$ gauge invariant. 
This can be done by dropping out the constraint (\ref{sci}).  
Altogether the independent constraints we choose are:
\bea
 \p_i \bp_i&=&0\label{conu1}\\
 \up_i-\bp_i&=&0\label{constr}
\eea
Using eqs.(\ref{conu1},\ref{constr})  we can write a first order Lagrangean equivalent to the initial one 
\be 
\ql' = \p_0 \ua_i \up_i + \p_0 \ba_i \bp_i - \of \up_i^2 -\of \bp_i^2 -
\hf H_i^2 + A_0 \p_i \bp_i + \Lambda_i ( \up_i-\bp_i).\label{la3}
\ee
Here, $\ua_0$, $\ba_0$, and $\Lambda_i$ are Lagrange multipliers and, as usual, $H_i=\hf\epsilon_{ijk}F_{jk}$.
Certainly, the canonical Hamiltonian of the model is not uniquely determined --- 
$\bH = \hf \up_i^2 - H_i^2$, 
$\bH' = \hf \bp_i^2 - H_i^2$ and so on are equally good and are a matter of redefinition of $\Lambda_i$. 

Eq.(\ref{conu1}) is the generator of the $U(1)$ gauge transformation of the field $\ba$, and eqs.(\ref{constr}) generate the  Stuckelberg symmetry.
We want to preserve the $U(1)$ symmetry but rid off from the  Stuckelberg one.
Therefore, we have to apply a particular gauge fixing.
(Note that $\ua_0$ and $\ba_0$ participate in (\ref{la3}) only through their combination  
$A_0=\ua_0+\ba_0$. 
This is a consequence of the already discussed fact that not all constraints are independent. 
Thus the combination of Lagrange multipliers $\hf(\ua_0+\ba_0)$ decouple from  all other fields and the functional integration over it is trivial.)
We want to interpret the conditions (\ref{zx1},\ref{zx2}) proposed in Refs.\cite{c2, c3} as gauge fixing conditions for the  Stuckelberg symmetry. 
Consider first eqs.(\ref{zx2}) ($\bF=0$). 
In general an antisymmetric rank $2$ tensor (which transforms  in  $(0,1)\oplus (1,0)$ representation of the Lorentz group) has  six independent components and same is the number of independent conditions in eqs.(\ref{zx2}).
But we have only three gauge generators and so, the closest to the conditions in eqs.(\ref{zx2}) we can use is
\be 
\bF_{jk} =0 \;\;\forall i, j \label{pg1}
\ee
or,  equivalently 
\be
\bar{H}_i=0. \label{pg2}
\ee
(It worth to be mentioned that eqs.(\ref{pg1}) are the  conditions used in the earliest work \cite{c1} but  thrown away in later works in favor of eqs.(\ref{zx2}).)
%

The great advantage of the gauge (\ref{pg2}) is that it does not fix the $U(1)$ freedom of the $\ba$ field (because the corresponding Poisson brackets with the $U(1)$ constraint (\ref{conu1}) are $0$).
However, eqs.(\ref{pg2}) do not fix the Stuckelberg gauge freedom either ---
the determinant of the matrix of Poisson brackets between constraints (\ref{constr}) and gauge fixing conditions (\ref{pg2}) is $0$.
A way out is to add a new gauge fixing condition and we have a  candidate ready for it --- eq.(\ref{zx1}).
It is easy to check that all third order minors of the rectangular  matrix of Poisson brackets between constraints (\ref{constr}) and gauge fixing conditions (\ref{zx1}, \ref{pg2}) are with non zero  (proportional to $\Delta$) determinant and thus we indeed have a gauge fixing.

The problem is that the gauge conditions are more than the  constraints and therefore one of them have to be dropped.
It is not possible this to be eq.(\ref{zx1}), so it must be one of the eqs.(\ref{pg2}).
But all of these conditions are independent which leads us to the conclusion that
the model in which we simultaneously fulfill eqs.(\ref{zx1},\ref{pg1}) is not equivalent to the QED.
The situation is even worse when we try to use eqs.(\ref{zx2}) as gauge fixing. 
In this case we have seven independent gauge condition  and only three gauge generators.

\section{The Electromagnetic field momentum}
Finally, some remarks on the definition of the conserved quantity momentum in QED which is in the heart of the discussion in Refs.\cite{c1, c2}.

The bare kinetic operator  corresponding to the Lagrangean (\ref{la1}) (after Fourier transformation) is
\be K = k^2\eta_{\mu \nu} - k_\mu k_\nu. \ee
Obviously it is non invertible, and
therefore it is not possible to make any  perturbative calculations in this theory.
The cure of the problem is to add into the Lagrangean a term proportional to $\p_\mu A^\mu$, e.g.
\be \delta L = - {\alpha\over 2} (\p_\mu A^\mu)^2. \ee
This is the so called $\alpha$ gauge term\footnote{
The usage of other Lorentz invariant gauge fixing terms requires generalization of the Noether theorem to Lagrangeans with higher derivatives.
}.
The kinetic operator now is 
$ K = k^2\eta_{\mu \nu} - (1-\alpha)k_\mu k_\nu$,
its inverse  is well defined, and so is the perturbation theory.
So,  we have to use not the Lagrangean $L$ from eq.(\ref{la1}) but $L+\delta L$ when we define the Noether  currents in the quantized theory. 
The choice $\alpha=1$ corresponds to the diagonal (Feynman - 't Hooht) gauge.
In this gauge the components of the gauge potential and the corresponding asymptotic states are solutions of the massless Klein--Gordon equation.  
This feature together with the requirement the operator $\p_\mu A^\mu$ between physical states to give $0$ solve the energy positiveness problem for the second quantized Electromagnetic field \cite{bs}.
When $\alpha\ne 1$  the kinetic operator is non diagonal but all its eigenvalues are proportional to $k^2$ ---
three of the them are  $k^2$ and one is $\alpha k^2$
(the  last eigenvalue corresponds to pure gauge degree of freedom).
The important point here is that for any $\alpha\ne 0$ the asymptotic states admit free particle interpretation.
However, the situation is completely different if there is no gauge fixing ($\alpha=0$) in which case the energy positiveness can not be proved.

The proposed in Refs.\cite{c1, c2, c3} representation of the Electromagnetic potential as sum of physical and pure gauge parts does not solve the problem with the kinetic operator kernel listed above.
The kinetic operators both for $\ua$ and $\ba$ fields have zero modes and some well chosen terms have to be added into the Lagrangean to improve the situation. 
The only possible term involving $\ua$ field is
$ 
\delta L' = f(\p_i \ua_i)
$
where $f$ is some function.
Unfortunately, this term is not Lorentz invariant and  plugging it into the Lagrangean results in  a non Lorentz invariant theory. 
This means  no plane wave external states and break down of the explicit Lorentz invariance in all orders of the perturbation theory. 
In other words --- a lot of problems.
Everything must be done from the scratch.
Probably, it is possible to be handled, as it is possible to deal with non invariant regularization, but the cleaver idea is to keep the manifest Lorentz invariance  throughout the calculations and fix the coordinate system at the final (the Coulomb gauge is equivalent to the Lorenz one plus fixation of the coordinate frame). See Refs.\cite{j1, j2} for more arguments on this point.

\section{Conclusions}
The performed analysis shows that the proposed decomposition of the Electromagnetic field into physical and pure gauge parts leads to a model which is not equivalent to the free Quantum Electrodynamics.
The Stuckelberg gauge symmetry which emerges in this decomposition is an Abelian symmetry and it acts only on the gauge field components.
Its generators are given by eqs.(\ref{constr}) and their form do not depend neither on the type of the gauge model in consideration (Abelian or non Abelian) nor on the interaction with additional matter fields.
Therefore, our analysis can be applied {\it en bloc} to any gauge system --- Abelian, non Abelian, without or with matter and the conclusion will be the same.

\section*{Acknowledgement} The work is supported by NSFB grant 2-288.

\end {document}